\documentclass[aps,preprint]{revtex4}%
\usepackage{amsfonts}
\usepackage{amsmath}
\usepackage{amssymb}
\usepackage{graphicx}%
\providecommand{\U}[1]{\protect\rule{.1in}{.1in}}

\begin{document}
\title{Dynamical Casimir effect for a massless scalar field between two concentric
spherical shells}
\author{F. Pascoal$^{1}$, L. C. C\'{e}leri$^{1,2}$, S. S. Mizrahi$^{1}$, and M. H. Y.
Moussa$^{3}$.}
\affiliation{$^{1}$Departamento de F\'{\i}sica, Universidade Federal de S\~{a}o Carlos,
Caixa Postal 676, S\~{a}o Carlos, 13565-905, S\~{a}o Paulo,\textit{ }Brazil}
\affiliation{$^{2}$Universidade Federal do ABC, Centro de Ci\^{e}ncias Naturais e Humanas,
R. Santa Ad\'{e}lia 166, Santo Andr\'{e}, 09210-170, S\~{a}o Paulo, Brasil.}
\affiliation{$^{3}$ Instituto de F\'{\i}sica de S\~{a}o Carlos, Universidade de S\~{a}o
Paulo, Caixa Postal 369, 13560-590 S\~{a}o Carlos, SP, Brazil }

\begin{abstract}
In this work we consider the dynamical Casimir effect for a massless scalar
field -- under Dirichlet boundary conditions -- between two concentric
spherical shells. We obtain a general expression for the average number of
particle creation, for an arbitrary law of radial motion of the spherical
shells, using two distinct methods: by computing the density operator of the
system and by calculating the Bogoliubov coefficients. We apply our general
expression to breathing modes: when only one of the shells oscillates and when
both shells oscillate in or out of phase. Since our results were obtained in
the framework of the perturbation theory, under resonant breathing modes they
are restricted to a short-time approximation.\textbf{ }We also analyze the
number of particle production and compare it with the results for the case of
plane geometry.

\end{abstract}

\pacs{PACS numbers:  31.30.jh; 03.65.-w; 03.70.+k; 42.50.Pq}
\maketitle

\section{Introduction}

The static Casimir effect, theoretically predicted in 1948 \cite{Casimir},
consists in the attraction of two perfectly conducting, parallel plates, due
to the distortion of the electromagnetic vacuum state. As a consequence of
quantum fluctuations under the very presence of external constraints, such a
distortion of the vacuum state emerges from the fundamental concepts of
quantum field theory \cite{Soff}. Casimir himself was the first to discuss the
importance of spherical geometry in the distortion of the vacuum state
\cite{Casimir56}, proposing in 1953 a semiclassical model for the stability of
the electron. In this model, the electron was assumed to be a perfectly
conducting spherical shell carrying a total charge $e$, with the Coulomb
repulsion balanced by an attractive Casimir force. However, as elucidated by
Boyer \cite{Boyer} in 1968, the Casimir force in spherical configuration is
repulsive, invalidating the attempt to explain the stability of the electron
through the Casimir force. Nonetheless, the electron model proposed by Casimir
provides evidence that one cannot predict whether the Casimir force will be
attractive or repulsive before the whole calculation is carried out. Moreover,
the development of the bag model of the hadrons in the early 1970s ---
describing hadrons as quarks and antiquarks confined inside of a spherical
cavity --- also stimulated the investigation of the Casimir effect with
spherical geometry \cite{Jhonson}.

Beyond static geometries, the dynamical Casimir effect (DCE), arising from
movable external constraints, probes even more deeply into the complexity of
the vacuum structure. The dynamics of the geometry gives rise to a
time-dependent Casimir force along with a dissipative component \cite{Farina}.
The mechanical energy dissipated by this `vacuum viscosity'\textbf{\ } induces
the most striking manifestation of the DCE, i.e., the mechanism of particle
creation and annihilation. G. T. Moore was the first to proceed to the
quantization of the radiation field in a cavity with movable perfectly
reflecting boundaries \cite{Moore} and, still in the 1970s, the creation of
photons from the nonuniform motion of the boundaries was predicted \cite{4,5}.
Later, it was also noticed that a sudden change in the refractive index of the
medium \cite{6,7,2} could also extract photons from the vacuum radiation field.

Regarding the spherical geometry --- the subject of the present work --- it
was proposed by Schwinger \cite{Sch} that the DCE could provide a possible
explanation for the sonoluminescence phenomenon, discovered in the 1930s
\cite{Sonoexp}. However, in spite of interesting results following Schwinger's
hypothesis \cite{Eberlein}, the theoretical description of sonoluminescence
still remains controversial . We finally mention that the DCE with spherical
geometry bears some similarity to the problem of particle creation in the
expanding Universe, where the spatial portion of the metric is a
hyperspherical surface with time-dependent radius
\cite{Parker,Carlitz,Davies,Fabio}.

In the present study we deal with the DCE for a massless scalar field confined
between two concentric spherical moving shells, and present a general
expression --- for any law of radial motion of the shells --- to compute the
average number of particle creation. We note that the particular case of a
unique spherical shell moving with a specific law of motion was studied in
Ref. \cite{Setare}. After deducing an effective Hamiltonian for the problem,
we compute the average number of particle creation by two distinct methods: by
considering the time evolution of the density operator of the cavity field and
also by computing the well-known Bogoliubov coefficients. Assuming, then, an
oscillatory radial motion for the spherical shells, our results are applied to
four different cases: when only (a) the inner or (b) the outer shell
oscillates, apart from those where both shells oscillate (c) in phase or (d)
out of phase.

The present paper is organized as follows: in section II we present the field
quantization and derive an effective Hamiltonian; in section III we calculate
the average number of particle production, for an arbitrary law of motion; in
section IV we specialize for four breathing modes and analyze our results;
finally, in section V the present our concluding remarks.

\section{Field quantization between two moving concentric spherical shells}

To quantize a massless scalar field between the two spherical shells, we start
from the action%

\begin{equation}
S=\int\operatorname*{d}t\operatorname*{d}\nolimits^{3}\mathbf{x\ }%
\mathcal{L}(\mathbf{x})=\frac{1}{2}\int\operatorname*{d}t\operatorname*{d}%
\nolimits^{3}\mathbf{x\ }\left(  \nabla\phi\cdot\nabla\phi-\frac{1}{c^{2}}%
\dot{\phi}^{2}\right)  \text{,} \label{1}%
\end{equation}
where the Lagrangian density $\mathcal{L}$ enables us to evaluate the
canonical momentum%
\begin{equation}
\pi(\mathbf{r},t)=\frac{\partial\mathcal{L}}{\partial\left(  \partial\dot
{\phi}\right)  }=\frac{1}{c^{2}}\dot{\phi}(\mathbf{r},t)\text{.} \label{2}%
\end{equation}
By minimizing the action (\ref{1}), we obtain the Klein-Gordon field equation
\begin{equation}
\left[  \frac{1}{c^{2}}\frac{\partial^{2}}{\partial t^{2}}-\nabla^{2}\right]
\phi(\mathbf{r},t)=0\text{,} \label{3}%
\end{equation}
where the cavity field $\phi(\mathbf{r},t)$ is subject to the Dirichlet
boundary conditions%
\begin{equation}
\phi(r\mathbf{=}r_{i},\theta,\varphi,t)=\phi(r=r_{o},\theta,\varphi
,t)=0\text{,} \label{4}%
\end{equation}
on the inner and outer spherical shells, with radii $r_{i}$ and $r_{o}$,
respectively. The spherical geometry of the cavity leads us, naturally, to
seek solutions for the cavity field and its canonical momentum in the form of
spherical harmonic expansions
\begin{subequations}
\label{4i}%
\begin{align}
\phi(\mathbf{r},t)  &  =\sum_{l=0}^{\infty}\sum_{m=-l}^{l}\sum_{s=1}^{\infty
}c\sqrt{\frac{\hslash}{2\omega_{ls}}}F_{ls}(r)\left[  c_{lms}(t)Y_{lm}%
(\theta,\varphi)+H.c\right]  ,\label{4ia}\\
\pi(\mathbf{r},t)  &  =-i\sum_{l=0}^{\infty}\sum_{m=-l}^{l}\sum_{s=1}^{\infty
}\frac{1}{c}\sqrt{\frac{\hslash\omega_{ls}}{2}}F_{ls}(r)\left[  c_{lms}%
(t)Y_{lm}(\theta,\varphi)-H.c\right]  . \label{4ib}%
\end{align}
By substituting the above expansions into Eqs. (\ref{3}) and (\ref{4}), we
obtain the differential equation
\end{subequations}
\begin{equation}
\frac{1}{r^{2}}\frac{\text{d}}{\text{d}r}\left(  r^{2}\frac{\text{d}F_{ls}%
(r)}{\text{d}r}\right)  +\left(  \frac{\omega_{ls}^{2}}{c^{2}}-\frac
{l(l+1)}{r^{2}}\right)  F_{ls}(r)=0, \label{5}%
\end{equation}
under the boundary conditions%
\[
F_{ls}(r=r_{i})=F_{ls}(r=r_{o})=0\text{.}%
\]
Moreover, as the radial functions are solutions to a boundary value problem,
they automatically satisfy the orthonormality relations%
\begin{equation}
\int_{r_{i}}^{r_{o}}F_{ls}(r)F_{ls^{\prime}}(r)r^{2}\ \text{d}r=\delta
_{s,s^{\prime}}\text{.} \label{6}%
\end{equation}

As the solution of Eq. (\ref{5}) is given by a linear combination of spherical
Bessel functions of the first ($j_{l}$) and second kind ($n_{l}$), the
boundary condition on the inner shell leads to the relation%

\[
F_{ls}(r)=\mathcal{N}_{ls}\left[  j_{l}\left(  \frac{\omega_{ls}r}{c}\right)
\ n_{l}\left(  \frac{\omega_{ls}r_{i}}{c}\right)  -j_{l}\left(  \frac
{\omega_{ls}r_{i}}{c}\right)  \ n_{l}\left(  \frac{\omega_{ls}r}{c}\right)
\right]  ,
\]
whereas that on the outer shell results in the transcendental equation
\begin{equation}
j_{l}\left(  \frac{\omega_{ls}r_{o}}{c}\right)  \ n_{l}\left(  \frac
{\omega_{ls}r_{i}}{c}\right)  -j_{l}\left(  \frac{\omega_{ls}r_{i}}{c}\right)
\ n_{l}\left(  \frac{\omega_{ls}r_{o}}{c}\right)  =0\text{.} \label{7}%
\end{equation}
The index $s$ in $\omega_{ls}$ --- which assumes discrete values and are not
necessarily equally spaced --- indicates the $s$th root of Eq. (\ref{7}). We
also note that the derivation of the solution of the problem of the moving
shells with dynamical boundary conditions follows directly from the
replacement of the static $r_{i(o)}$ by the dynamical radii $r_{i(o)}(t)$,
since all time dependence in the system arises from them. In Fig. 1 we have
constructed a map of the solutions of the Eq. (\ref{7}) for some values of the
numbers $l$ and $s$. As we can see, for the case $l=0$, the frequencies are
equidistant, which does not occur for the case $l\neq0$. However, if we have
$r_{o}\left(  t\right)  >>r_{o}\left(  t\right)  -r_{i}$ $\left(  t\right)  $
(i. e., when the radii of the shells are much larger than the separation
between them), the solutions for all values of $l$ approach to the solution
for $l=0$, i.e., $\omega_{ls}\rightarrow\omega_{0s}$.

\subsection{Canonical field quantization}

The canonical quantization of the scalar field $\phi$ in Eq. (\ref{4ia}) is
performed --- by promoting the coefficients $c_{lms}$ and $c_{lms}^{\ast}$ to
operators $a_{lms}$ and $a_{l^{\prime}m^{\prime}s^{\prime}}^{\dagger}$ ---
through the construction of a field operator $\hat{\phi}$ satisfying Eqs.
(\ref{3})-(\ref{4}) and the equal-time commutation relation
\begin{subequations}
\label{8}%
\begin{align*}
\left[  \hat{\phi}(\mathbf{r},t),\hat{\pi}(\mathbf{r}^{\prime},t)\right]   &
=i\hslash\delta^{3}(\mathbf{r}-\mathbf{r}^{\prime}),\\
\left[  \hat{\phi}(\mathbf{r},t),\hat{\phi}(\mathbf{r}^{\prime},t)\right]   &
=\left[  \hat{\pi}(\mathbf{r},t),\hat{\pi}(\mathbf{r}^{\prime},t)\right]
=0\text{,}%
\end{align*}
where $\hat{\pi}$ is the momentum operator associated with $\pi$. The above
relations between the field operators automatically imply the bosonic
commutation relations for the standard creation and annihilation operators
\end{subequations}
\begin{align*}
\left[  a_{lms}(t),a_{l^{\prime}m^{\prime}s^{\prime}}^{\dagger}(t)\right]   &
=\delta_{ll^{\prime}}\delta_{mm^{\prime}}\delta_{ss^{\prime}},\\
\left[  a_{lms}(t),a_{l^{\prime}m^{\prime}s^{\prime}}(t)\right]   &  =\left[
a_{lms}^{\dagger}(t),a_{l^{\prime}m^{\prime}s^{\prime}}^{\dagger}(t)\right]
=0.
\end{align*}
Through the time derivative of the quantum version of Eqs. (\ref{4i}),
together with the equations for the cavity field (\ref{3}) and its canonical
momentum (\ref{2}), we obtain the Heisenberg equation for the annihilation
operators
\begin{equation}
\dot{a}_{lms}(t)=-i\omega_{ls}(t)a_{lms}(t)+\sum_{s^{\prime}}\mu_{l\left[
ss^{\prime}\right]  }(t)a_{lms^{\prime}}(t)+\sum_{s^{\prime}}\mu_{l\left(
ss^{\prime}\right)  }(t)a_{l(-m)s^{\prime}}^{\dag}(t), \label{8i}%
\end{equation}
where $\mu_{l\left(  ss^{\prime}\right)  }\left(  t\right)  =\left[
\mu_{lss^{\prime}}(t)+\mu_{ls^{\prime}s}(t)\right]  /2$ and $\mu_{l\left[
ss^{\prime}\right]  }\left(  t\right)  =\left[  \mu_{lss^{\prime}}%
(t)-\mu_{ls^{\prime}s}(t)\right]  /2$ are the symmetric and antisymmetric
parts, respectively, of the coefficient
\[
\mu_{lss^{\prime}}(t)=\frac{\dot{\omega}_{ls}(t)}{2\omega_{ls}(t)}%
\delta_{ss^{\prime}}+\left(  1-\delta_{ss^{\prime}}\right)  \sqrt{\frac
{\omega_{ls}(t)}{\omega_{ls^{\prime}}(t)}}\int_{r_{i}(t)}^{r_{o}(t)}%
r^{2}F_{ls^{\prime}}(r;t)\dot{F}_{ls}(r;t)\operatorname*{d}r.
\]
From Eq. (\ref{8i}) we directly obtain $\dot{a}_{lms}^{\dagger}$.

\subsection{An effective Hamiltonian}

Following the reasoning exposed in Ref. \cite{Law}, we next derive an
effective Hamiltonian governing the evolution of the creation and annihilation
operators, as given by Eq. (\ref{8i}). To this end, we consider the most
general form of a quadratic Hamiltonian
\begin{align*}
H_{eff}  &  =\hslash\sum_{l,l^{\prime}}\sum_{m,m^{\prime}}\sum_{s,s^{\prime}%
}\left[  f_{ll^{\prime}mm^{\prime}ss^{\prime}}^{(1)}(t)a_{lms}^{\dag
}a_{l^{\prime}m^{\prime}s^{\prime}}^{\dagger}+f_{ll^{\prime}mm^{\prime
}ss^{\prime}}^{(2)}(t)a_{lms}^{\dag}a_{l^{\prime}m^{\prime}s^{\prime}}\right.
\\
&  \left.  +f_{ll^{\prime}mm^{\prime}ss^{\prime}}^{(3)}(t)a_{lms}a_{l^{\prime
}m^{\prime}s^{\prime}}^{\dagger}+f_{ll^{\prime}mm^{\prime}ss^{\prime}}%
^{(4)}(t)a_{lms}a_{l^{\prime}m^{\prime}s^{\prime}}\right]  ,
\end{align*}
which governs the evolution $\dot{a}_{lms}=\left(  i/\hslash\right)  \left[
H_{eff},a_{lms}\right]  $. Comparing the evolution equations for $\dot
{a}_{lms}(t)$ and $\dot{a}_{lms}^{\dagger}(t)$ obtained through the Heisenberg
equation of motion with those following from Eq. (\ref{8i}), we obtain the
effective Hamiltonian $H_{eff}(t)=H_{0}(t)+V(t)$, where
\begin{subequations}
\begin{align}
H_{0}(t)  &  =\hslash\sum_{l,m,s}\omega_{ls}(t)\left(  a_{lms}^{\dag}%
a_{lms}+\frac{1}{2}\right)  ,\label{9a}\\
V(t)  &  =i\frac{\hslash}{2}\sum_{l,m}\sum_{s,s^{\prime}}\mu_{lss^{\prime}%
}(t)\left[  \left(  a_{lms^{\prime}}+a_{l(-m)s^{\prime}}^{\dag}\right)
a_{lms}^{\dag}-a_{lms}\left(  a_{l\left(  -m\right)  s^{\prime}}%
+a_{lms^{\prime}}^{\dag}\right)  \right]  \text{.} \label{9b}%
\end{align}

In what follows we compute the average number of particles created in a
selected mode via two distinct methods: the density operator and the
Bogoliubov coefficients.

\section{Average number of particle creation}

\subsection{The density operator}

From our effective Hamiltonian $H_{eff}(t)$ we obtain, in the interaction
picture, the density operator of the cavity modes
\end{subequations}
\[
\rho(t)=\rho(0)+\sum_{n=1}^{\infty}\left(  -\frac{i}{\hslash}\right)  ^{n}%
\int_{0}^{t}\operatorname*{d}t_{1}\int_{0}^{t_{1}}\operatorname*{d}t_{2}%
\cdot\cdot\cdot\int_{0}^{t_{n-1}}\operatorname*{d}t_{n}\left[  V_{I}%
(t_{1}),\left[  V_{I}(t_{2}),\cdot\cdot\cdot\left[  V_{I}(t_{n}),\rho
(0)\right]  \right]  \right]
\]
where%
\[
V_{I}(t)=i\frac{\hslash}{2}\sum_{l,m}\sum_{s,s^{\prime}}\mu_{lss^{\prime}%
}(t)\left[  \left(  \tilde{a}_{lms^{\prime}}(t)+\tilde{a}_{l(-m)s^{\prime}%
}^{\dag}(t)\right)  \tilde{a}_{lms}^{\dag}(t)-\tilde{a}_{lms}(t)\left(
\tilde{a}_{l(-m)s^{\prime}}(t)+\tilde{a}_{lms^{\prime}}^{\dag}(t)\right)
\right]
\]
with $\tilde{a}_{lms}(t)=a_{lms}\exp\left(  -i\Omega_{ls}(t)\right)  $,
$\tilde{a}_{lms}^{\dag}(t)=a_{lms}^{\dag}\exp\left(  i\Omega_{ls}(t)\right)
$, and $\Omega_{ls}(t)=\int_{0}^{t}$d$t_{1}\ \omega_{ls}(t_{1})$.

To compute the average number of particles created in a particular mode
labeled by the quantum numbers ($l,m,s$), given by $N_{lms}%
(t)=\operatorname*{Tr}\rho(t)a_{lms}^{\dag}a_{lms}$, we go up to the
second-order approximation in the velocity of the cavity walls, $\dot{r}_{i}$,
$\dot{r}_{0}\ll c$. Starting from the vacuum state $\rho(0)=\left\vert
\left\{  0\right\}  \right\rangle \left\langle \left\{  0\right\}  \right\vert
$, we thus obtain%

\begin{align}
N_{lms}(t)  &  \simeq\frac{1}{4}\int_{0}^{t}\operatorname*{d}t_{1}\int
_{0}^{t_{1}}\operatorname*{d}t_{2}\sum_{m^{\prime},m^{\prime\prime}}%
\sum_{q,q^{\prime}}\sum_{p,p^{\prime}}\mu_{lpp^{\prime}}(t_{1})\mu
_{lqq^{\prime}}(t_{2})\nonumber\\
&  \times\left\{  \left\langle \left\{  0\right\}  \right\vert a_{lm^{\prime
\prime}q}a_{l(-m^{\prime\prime})q^{\prime}}a_{lms}^{\dag}a_{lms}%
a_{l(-m^{\prime})p^{\prime}}^{\dag}a_{lm^{\prime}p}^{\dag}\left\vert \left\{
0\right\}  \right\rangle \right. \nonumber\\
&  \times\exp\left[  -i\left(  \Omega_{lp}(t_{1})+\Omega_{lp^{\prime}}%
(t_{1})-\Omega_{lq}(t_{2})-\Omega_{lq^{\prime}}(t_{2})\right)  \right]
\nonumber\\
&  +\left\langle \left\{  0\right\}  \right\vert a_{lm^{\prime}p}%
a_{l(-m^{\prime})p^{\prime}}a_{lms}^{\dag}a_{lms}a_{l(-m^{\prime\prime
})q^{\prime}}^{\dag}a_{lm^{\prime\prime}q}^{\dag}\left\vert \left\{
0\right\}  \right\rangle \nonumber\\
&  \left.  \times\exp\left[  i\left(  \Omega_{lp}(t_{1})+\Omega_{lp^{\prime}%
}(t_{1})-\Omega_{lq}(t_{2})-\Omega_{lq^{\prime}}(t_{2})\right)  \right]
\right\}  . \label{10}%
\end{align}
Using the result%
\begin{align*}
\left\langle \left\{  0\right\}  \right\vert a_{lm^{\prime\prime}%
q}a_{l(-m^{\prime\prime})q^{\prime}}a_{lms}^{\dag}a_{lms}a_{l(-m^{\prime
})p^{\prime}}^{\dag}a_{lm^{\prime}p}^{\dag}\left\vert \left\{  0\right\}
\right\rangle  &  =\delta_{m^{\prime\prime},m}\delta_{m^{\prime},m}%
\delta_{q^{\prime},p_{l}^{\prime}}\delta_{s,p}\delta_{s_{l},q_{l}}%
+\delta_{m^{\prime\prime},m}\delta_{-m^{\prime},m}\delta_{q^{\prime},p}%
\delta_{s,p^{\prime}}\delta_{s,q}\\
&  +\delta_{-m^{\prime\prime},m}\delta_{m^{\prime},m}\delta_{q,p^{\prime}%
}\delta_{s,p}\delta_{s,q^{\prime}}+\delta_{-m^{\prime\prime},m}\delta
_{-m^{\prime},m}\delta_{q,p}\delta_{s,p^{\prime}}\delta_{s,q^{\prime}}\text{,}%
\end{align*}
the expression (\ref{10}) reduces to the compact form
\begin{equation}
N_{lms}(t)=\sum_{s^{\prime}}\left\vert \int_{0}^{t}\text{d}t_{1}\mu_{l\left(
s^{\prime}s\right)  }(t_{1})\exp\left\{  i\left[  \Omega_{ls^{\prime}}%
(t_{1})+\Omega_{ls}(t_{1})\right]  \right\}  \right\vert ^{2}, \label{11}%
\end{equation}
which, like the energy of a given mode $\omega_{ls}$, is the same for any
value of $m$.

\subsection{The Bogoliubov coefficients}

In this section we compute the average number of particle creation by means of
the Bogoliubov coefficients \cite{Bogoliubov}, defined as%

\begin{subequations}
\label{12}%
\begin{align}
a_{lms}(t)  &  =\sum_{q=1}^{\infty}\alpha_{lsq}(t)a_{lmq}(0)+\sum
_{q=1}^{\infty}\beta_{lsq}(t)a_{l\left(  -m\right)  q}^{\dag}(0),\label{12a}\\
a_{lms}^{\dag}(t)  &  =\sum_{q=1}^{\infty}\alpha_{lsq}^{\ast}(t)a_{lmq}^{\dag
}(0)+\sum_{q=1}^{\infty}\beta_{lsq}^{\ast}(t)a_{l\left(  -m\right)  q}(0),
\label{12b}%
\end{align}
with the initial conditions $\alpha_{lsq}(0)=\delta_{s,q}$ and $\beta
_{lsq}(0)=0$. In what follows, we obtain the time derivatives $\dot{a}%
_{lms}(t)$ and $\dot{a}_{lms}^{\dagger}(t)$ directly from Eq. (\ref{12}):
\end{subequations}
\begin{subequations}
\label{13}%
\begin{align}
\dot{a}_{lms}(t)  &  =\sum_{q}\dot{\alpha}_{lsq}(t)a_{lmq}(0)+\sum_{q}%
\dot{\beta}_{lsq}(t)a_{l\left(  -m\right)  q}^{\dag}(0),\label{13a}\\
\dot{a}_{lms}^{\dag}(t)  &  =\sum_{q}\dot{\alpha}_{lsq}^{\ast}(t)a_{lmq}%
^{\dag}(0)+\sum_{q}\dot{\beta}_{lsq}^{\ast}(t)a_{l\left(  -m\right)  q}(0),
\label{13b}%
\end{align}
and also by substituting $a_{lms}(t)$ and $a_{lms}^{\dagger}(t)$ from Eq.
(\ref{12}) into Eq. (\ref{8}), we get%

\end{subequations}
\begin{subequations}
\label{14}%
\begin{align}
\dot{a}_{lms}(t)  &  =\sum_{s^{\prime},q}\left[  -i\omega_{ls}\delta
_{s,s^{\prime}}\alpha_{lsq}(t)+\mu_{l\left[  ss^{\prime}\right]  }%
\alpha_{ls^{\prime}q}(t)+\mu_{l\left(  ss^{\prime}\right)  }\beta_{ls^{\prime
}q}^{\ast}(t)\right]  a_{lmq}(0)\nonumber\\
&  +\sum_{s^{\prime},q}\left[  -i\omega_{ls}\delta_{s,s^{\prime}}\beta
_{lsq}(t)+\mu_{l\left[  ss^{\prime}\right]  }\beta_{ls^{\prime}q}%
(t)+\mu_{l\left(  ss^{\prime}\right)  }\alpha_{ls^{\prime}q}^{\ast}(t)\right]
a_{l\left(  -m\right)  q}^{\dag}(0),\label{14a}\\
\dot{a}_{lms}^{\dag}(t)  &  =\sum_{s^{\prime},q}\left[  i\omega_{ls}%
\delta_{s,s^{\prime}}\alpha_{lsq}^{\ast}(t)+\mu_{l\left[  ss^{\prime}\right]
}\alpha_{ls^{\prime}q}^{\ast}(t)+\mu_{l\left(  ss^{\prime}\right)  }%
\beta_{ls^{\prime}q}(t)\right]  a_{lmq}^{\dag}(0)\nonumber\\
&  +\sum_{s^{\prime},q}\left[  i\omega_{ls}\delta_{s,s^{\prime}}\beta
_{lsq}^{\ast}(t)+\mu_{l\left[  ss^{\prime}\right]  }\beta_{ls^{\prime}q}%
^{\ast}(t)+\mu_{l\left(  ss^{\prime}\right)  }\alpha_{ls^{\prime}q}(t)\right]
a_{l\left(  -m\right)  q}(0). \label{14b}%
\end{align}
By equating terms in the expressions for $\dot{a}_{lms}(t)$ and $\dot{a}%
_{lms}^{\dagger}(t)$ in Eqs. (\ref{13}) and (\ref{14}) we obtain, after some
algebraic manipulation, two coupled differential equations for the Bogoliubov coefficients%

\end{subequations}
\begin{subequations}
\label{15}%
\begin{align}
\dot{\alpha}_{lss^{\prime}}(t)  &  =-i\omega_{ls}(t)\alpha_{lss^{\prime}%
}(t)+\sum_{q}\left[  \mu_{l\left[  sq\right]  }(t)\alpha_{lqs^{\prime}}%
(t)+\mu_{l\left(  sq\right)  }(t)\beta_{lqs^{\prime}}^{\ast}(t)\right]
,\label{15a}\\
\dot{\beta}_{lss^{\prime}}(t)  &  =-i\omega_{ls}(t)\beta_{lss^{\prime}}%
(t)\sum_{q}\left[  \mu_{l\left[  sq\right]  }(t)\beta_{lqs^{\prime}}%
(t)+\mu_{l\left(  sq\right)  }(t)\alpha_{lqs^{\prime}}^{\ast}(t)\right]  .
\label{15b}%
\end{align}
Now, we expand these coefficients in powers of the coupling strength
$\mu_{lss^{\prime}}$, such that
\end{subequations}
\begin{subequations}
\label{16}%
\begin{align}
\alpha_{lss^{\prime}}(t)  &  =\operatorname*{e}\nolimits^{-i\Omega_{ls}%
(t)}\sum_{\lambda=0}^{\infty}\alpha_{lss^{\prime}}^{\left(  \lambda\right)
}(t)\label{16a}\\
\beta_{lss^{\prime}}(t)  &  =\operatorname*{e}\nolimits^{-i\Omega_{ls}(t)}%
\sum_{\lambda=0}^{\infty}\beta_{lss^{\prime}}^{\left(  \lambda\right)
}(t)\text{,} \label{16b}%
\end{align}
where the factor $\operatorname*{e}\nolimits^{-i\Omega_{ls}(t)}$ is introduced
for convenience. By substituting Eq. (\ref{16}) into Eq. (\ref{15}) we derive
the recurrence relations
\end{subequations}
\begin{subequations}
\label{17}%
\begin{align}
\alpha_{lss^{\prime}}^{\left(  \lambda\right)  }(t)  &  =\int_{0}^{t}%
\text{d}t_{1}\operatorname*{e}\nolimits^{i\Omega_{ls}(t_{1})}\sum_{q}\left[
\mu_{l\left[  sq\right]  }(t_{1})\operatorname*{e}\nolimits^{-i\Omega
_{lq}(t_{1})}\alpha_{lqs^{\prime}}^{\left(  \lambda-1\right)  }(t_{1}%
)+\mu_{l\left(  sq\right)  }(t_{1})\operatorname*{e}\nolimits^{i\Omega
_{lq}(t_{1})}\beta_{lqs^{\prime}}^{\left(  \lambda-1\right)  \ast}%
(t_{1})\right]  ,\label{17a}\\
\beta_{lss^{\prime}}^{\left(  \lambda\right)  }(t)  &  =\int_{0}^{t}%
\text{d}t_{1}\operatorname*{e}\nolimits^{i\Omega_{ls}(t_{1})}\sum_{q}\left[
\mu_{l\left[  sq\right]  }(t_{1})\operatorname*{e}\nolimits^{-i\Omega
_{lq}(t_{1})}\beta_{lqs^{\prime}}^{\left(  \lambda-1\right)  }(t_{1}%
)+\mu_{l\left(  sq\right)  }(t_{1})\operatorname*{e}\nolimits^{i\Omega
_{lq}(t_{1})}\alpha_{lqs^{\prime}}^{\left(  \lambda-1\right)  \ast}%
(t_{1})\right]  . \label{17b}%
\end{align}
When the initial conditions, given in the zeroth-order terms $\alpha
_{lss^{\prime}}^{(0)}(t)=\delta_{s,s^{\prime}}$ and $\beta_{lss^{\prime}%
}^{(0)}(t)=0$, are substituted back into Eq. (\ref{17b}), we finally obtain
the first-order solution%

\end{subequations}
\begin{equation}
\beta_{lss^{\prime}}^{\left(  1\right)  }(t)=\int_{0}^{t}\operatorname*{d}%
t_{1}\operatorname*{e}\nolimits^{i\left[  \Omega_{ls}(t_{1})+\Omega
_{ls^{\prime}}(t_{1})\right]  }\mu_{l\left(  ss^{\prime}\right)  }%
(t_{1})\text{.} \label{18}%
\end{equation}

We next compute the average number of particle creation from the expression
$N_{lms}(t)=\left\langle \left\{  0\right\}  \right\vert a_{lms}^{\dag
}(t)a_{lms}(t)\left\vert \left\{  0\right\}  \right\rangle =\sum_{s^{\prime}%
}\left\vert \beta_{l,s,s^{\prime}}(t)\right\vert ^{2}$, where $\left\vert
\left\{  0\right\}  \right\rangle $ indicates the initial vacuum state of the
cavity and $t$ is the time interval during which the shells have been in
motion. Up to second order in the coupling coefficients $\mu_{lss^{\prime}}$,
we do obtain the result
\begin{equation}
N_{lms}(t)\simeq\sum_{s^{\prime}}\left\vert \beta_{lss^{\prime}}^{\left(
1\right)  }(t)\right\vert ^{2}\text{.} \label{19}%
\end{equation}
The substitution of Eq. (\ref{18}) into Eq. (\ref{19}) finally gives exactly
the same Eq. (\ref{11}).

\section{Particle creation between harmonically oscillating shells}

In this section, we assume that the shells perform small harmonic oscillations
described by
\begin{equation}
r_{\alpha}(t)=r_{\alpha}\left[  1+\epsilon_{\alpha}\sin\left(  \varpi
t\right)  \right]  ,\text{ }\alpha=i,o\text{.} \label{A1}%
\end{equation}
where $\epsilon_{\alpha}\ll1$ and $\varpi$ stands for the frequency associated
with the oscillating shells. By substituting Eq. (\ref{A1}) into Eq.
(\ref{11}) we obtain, up to second order in $\epsilon_{\alpha}$, the result%

\begin{align}
N_{lms}  &  \simeq\sum_{s^{\prime}}\left\vert \left(  \frac{\exp\left[
i\left(  \omega_{lss^{\prime}}+\varpi\right)  t\right]  -1}{\left(
\omega_{lss^{\prime}}+\varpi\right)  }+\frac{\exp\left[  i\left(
\omega_{lss^{\prime}}-\varpi\right)  t\right]  -1}{\left(  \omega
_{lss^{\prime}}-\varpi\right)  }\right)  \right\vert ^{2}\nonumber\\
&  \times\left(  \sum_{\alpha}c_{l\left(  ss^{\prime}\right)  }^{\alpha
}r_{\alpha}\epsilon_{\alpha}\varpi\right)  ^{2}, \label{A2}%
\end{align}
where we have defined%
\begin{align}
c_{lss^{\prime}}^{\alpha}  &  \equiv\frac{1}{2\omega_{ls}(0)}\frac
{\partial\omega_{ls}(0)}{\partial r_{\alpha}}\delta_{ss^{\prime}}+\left(
1-\delta_{ss^{\prime}}\right)  \sqrt{\frac{\omega_{ls}(0)}{\omega_{ls^{\prime
}}(0)}}\int_{r_{i}}^{r_{o}}r^{2}F_{ls^{\prime}}(r;0)\frac{\partial
F_{ls}(r;0)}{\partial r_{\alpha}}\operatorname*{d}r\nonumber\\
&  +\left(  1-\delta_{ss^{\prime}}\right)  \sqrt{\frac{\omega_{ls}(0)}%
{\omega_{ls^{\prime}}(0)}}\frac{\partial\omega_{ls}(0)}{\partial r_{\alpha}%
}\int_{r_{i}}^{r_{o}}r^{2}F_{ls^{\prime}}(r;0)\frac{\partial F_{ls}%
(r;0)}{\partial\omega_{ls}(0)}\operatorname*{d}r, \label{A3}%
\end{align}
and $\omega_{lss^{\prime}}\equiv\omega_{ls}(0)+\omega_{ls^{\prime}}(0)$. From
Eq. (\ref{A2}), we observe the occurrence of resonances when $\varpi
=\omega_{lss^{\prime}}$, and for a given mode $l,s$, the average particle
creation in the $s^{\prime}$th resonance is given by
\begin{equation}
\lim_{\varpi\rightarrow\omega_{lss^{\prime}}}N_{lms}(t)\simeq\left(
\sum_{\alpha}c_{l\left(  ss^{\prime}\right)  }^{\alpha}r_{\alpha}%
\epsilon_{\alpha}\varpi t\right)  ^{2} \label{A31}%
\end{equation}
exhibiting a quadratic increase with time.

For $l=0$ the transcendental equation (\ref{7}), has the analytical solution
\[
\omega_{0s}(t)=s\omega_{01}(t)=\frac{s\pi c}{r_{0}(t)-r_{i}(t)},
\]
relating the radii to the instantaneous field frequencies for mode $s$,
implying the resonance condition $\varpi=\left(  s+s^{\prime}\right)
\omega_{01}(0)$. The coefficients (\ref{A3}) thus reduces to
\[
c_{0\left(  ss^{\prime}\right)  }^{i}=-\left(  -1\right)  ^{s+s^{\prime}%
}c_{0\left(  ss^{\prime}\right)  }^{o}=\frac{\sqrt{ss^{\prime}}}{s+s^{\prime}%
}\frac{1}{r_{o}-r_{i}}\text{,}%
\]
and the average number of particle creation is given by
\begin{equation}
\lim_{\varpi\rightarrow\omega_{0ss^{\prime}}}N_{0ms}(t)\simeq\left(
\frac{ss^{\prime}}{\left(  s+s^{\prime}\right)  ^{2}}\right)  \left(
\frac{\epsilon_{o}r_{o}-\left(  -1\right)  ^{s+s^{\prime}}\epsilon_{i}r_{i}%
}{r_{o}-r_{i}}\right)  ^{2}\left(  \varpi t\right)  ^{2}\text{.} \label{A4}%
\end{equation}
Notice that since the lower bound of $r_{o}-r_{i}$ is $\left\vert
r_{o}\epsilon_{o}\right\vert +\left\vert r_{i}\epsilon_{i}\right\vert $, the
maximum value of the second factor on the RHS of Eq. (\ref{A4}) is $1$. This
is in agreement with the fact that the Casimir effect is more pronounced at
small distances between the shells. Moreover, the effective velocities
$\varpi\epsilon_{o}r_{o}-\left(  -1\right)  ^{s+s^{\prime}}\varpi\epsilon
_{i}r_{i}=v_{o}-\left(  -1\right)  ^{s+s^{\prime}}v_{i}$ play an important
role on the particle creation process, which is compatible with the results of
plane geometry \cite{velocity}.

For the case $l\neq0$, the resonances are shifted to noninteger values of the
ratio $\varpi/\omega_{01}(0)$, since the eigenfrequencies $\omega_{ls}$ are no
longer equidistant, as can be seen in Fig. 1. In this case Eq. (\ref{7}) has
no analytical solution, so we can not write an obvious closed expression for
the coefficients $c_{lss^{\prime}}^{\alpha}$. In Fig.2 we plot $\left(
r_{o}-r_{i}\right)  \left\vert c_{lss^{\prime}}^{\alpha}\right\vert $ as a
function of the ratio $r_{o}/r_{i}$. As we can see, the cases $l=0$ and
$l\neq0$ exhibits the same behavior, i. e.,\ as the distance between the
shells decreases, $\left\vert c_{lss^{\prime}}^{\alpha}\right\vert $
increases. In contrast, when $l\neq0$, where $c_{l(ss^{\prime})}^{i}%
\neq-(-1)^{s+s^{\prime}}c_{l(ss^{\prime})}^{o}$, the effective velocities
$v_{o}-\left(  -1\right)  ^{s+s^{\prime}}v_{i}$ do not exhibits an evident
role in the amplitude of the coefficients $c_{l(ss^{\prime})}^{\alpha}$. This
fact shows that the qualitative difference between the plane and spherical
geometries appears essentially for $l\neq0$, as evidenced from Fig. 1.

Under the law of motion (\ref{A1}), we analyze the average number of particle
creation in the four different cases mentioned in the introduction: when (a)
only the inner shell oscillates ($\epsilon_{i}=\epsilon$ and $\epsilon_{o}%
=0$); (b) only the outer shell oscillates ($\epsilon_{i}=0$ and $\epsilon
_{o}=\epsilon$); (c) both shells oscillate in phase ($\epsilon_{i}%
=\epsilon_{o}=\epsilon$); and (d) both shells oscillate out of phase
($\epsilon_{i}=-\epsilon_{o}=\epsilon$).

In Fig. 3 we present the plot of the ratio $N_{lms}(t)/(\epsilon\varpi t)^{2}$
against $\varpi/\omega_{01}(0)$, under the resonance condition ($\varpi
=\omega_{lss^{\prime}}$), for all four cases and for few values of $l$ and
$s$. We find that the principal resonance --- which maximizes $N_{lms}(t)$ ---
occurs when $\varpi=2s\omega_{lss}(0)$ ($s^{\prime}=s$) in cases (a), (b), an
(c), as expected. However, in case (d), this resonance can be shifted for the
value $s^{\prime}=s+1$, depending on the ratio $r_{o}/r_{i}$ which, for $l=0$
reads%
\[
\left\vert \frac{v_{o}+v_{i}}{v_{o}-v_{i}}\right\vert >\sqrt{1+\frac
{1}{4s\left(  s+1\right)  }}\text{.}%
\]
This result follows directly from Eq. (\ref{A4}). Note that the case where
only the outer shell oscillates produces a larger number of particles than
that where only the inner one oscillates. This can be observed directly from
Eq. (\ref{A31}) under the assumption of only one oscillating shell, where the
rate $N_{lms}(t)$ is proportional to the velocity of the moving shell. We also
note from Fig. 3 that as $l$ increases, the particle creation rate in this
mode decreases accordingly. This result is expected, since the energy of a
given mode increases with $l$. By its turn, for the cases (c) and (d) we can
have a larger or a smaller number of particle created in the cavity, depending
on the chosen resonance: if $s+s^{\prime}$ is an even number, the case (c)
will present a smaller number of particles than that all other cases while the
case (d) will present the larger number of particles among the four cases. If
$s+s^{\prime}$ is an odd number, the opposite situation will occur.

In Eq. (\ref{A31}) we also observe that no particle can be created, even under
the resonance condition, under the following constraint%
\[
\frac{r_{o}}{r_{i}}=-\frac{\epsilon_{i}c_{l\left(  ss^{\prime}\right)  }^{i}%
}{\epsilon_{o}c_{l\left(  ss^{\prime}\right)  }^{o}}>1\text{,}%
\]
which for $l=0$ reduces to%
\[
\frac{r_{o}}{r_{i}}=\frac{\epsilon_{i}}{\epsilon_{o}}\left(  -1\right)
^{s+s^{\prime}}>1\text{,}%
\]
showing that, if $s+s^{\prime}$ is an even (odd) number and $\epsilon
_{i}/\epsilon_{o}>0$ ($<0$) this condition is not satisfied and we will always
have particles created in the cavity.

We stress that since the expression for the average number of particle
creation was obtained in the frameworks of the perturbation theory, for
resonant breathing modes it is valid only in the short-time approximation,
$\epsilon_{\alpha}\varpi t\ll1$. Therefore, the analysis performed above
cannot predict the real number of the created particles in the long time limit.

\section{Concluding remarks}

Here we have considered two concentric spherical shells that are allowed to
move, and we analyzed the particle creation within the DCE for the massless
scalar field confined in the cavity. The particle creation were computed for
an arbitrary law of radial motion of the spherical shells, using two distinct
methods: the density operator of the system and the Bogoliubov coefficients.
We applied our general results to the case of an oscillatory radial motion of
the spherical shells, associated with breathing modes, identifying the
resonance conditions were the number of particle creation is more significant.
Analyzing these resonances, we have noted that qualitative differences between
the plane and spherical geometries arise when $l\neq0$. We have considered
four distinct cases of the breathing modes: when only (a) the inner or (b) the
outer shell oscillates, or both shells oscillate (c) in phase or (d) out of
phase. As already emphasized, our resonant results are restricted to the
short-time approximation $\epsilon_{\alpha}\varpi t\ll1$. In conclusion,we
believe that the present work is enlarging perspectives in the subject of the
DCE, which is getting the interest of both,theoreticians and experimentalists.

\begin{acknowledgments}
Acknowledgements
\end{acknowledgments}

We wish to express thanks for the support from FAPESP and CNPq, Brazilian agencies.

\textbf{Figure captions}

Fig. 1 (color online). Map of the solutions of the transcendental equation
(\cite{7}). The colors correspond to different values of the number $l$: the
black lines are for $l=0$, the red ones for $l=1$, and for $l=2$ the blue
lines. The solid, dashed, and dotted lines correspond to the cases $s=1$,
$s=2$, and $s=3$, respectively.

Fig. 2 (color online). Plot of $\left(  r_{o}-r_{i}\right)  \left\vert
c_{l1s^{\prime}}^{\alpha}\right\vert $ against the ratio $r_{o}/r_{i}$. The
solid and dashed lines correspond to $s^{\prime}=1$ and $s^{\prime}=2$,
respectively. The black line correspond to $l=0$ and $\alpha=i,o$. The blue
and red lines are for $\alpha=i$ and $\alpha=o$, respectively, both with $l=1$.

Fig. 3 Plot of $N_{lms}(t)/(\epsilon\varpi t)$ against $\varpi/\omega_{01}%
(0)$, in the resonance condition, for the cases (a), (b), (c), and (d) for few
values of $l$ and $s$. We have set $r_{o}=2r_{i}$.

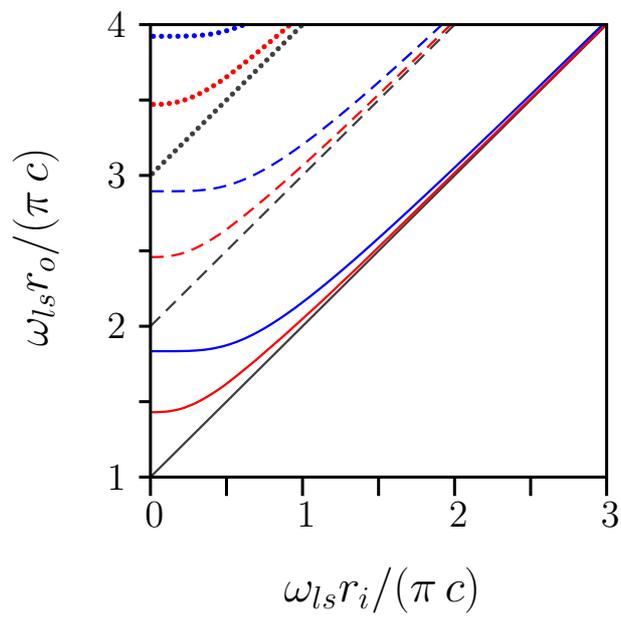
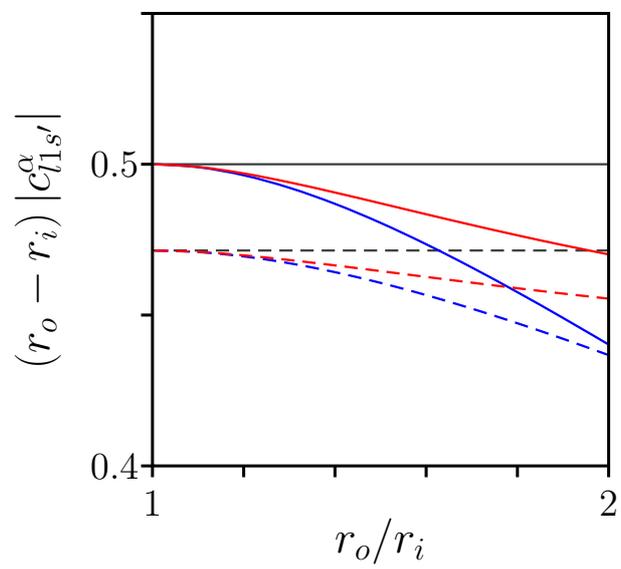
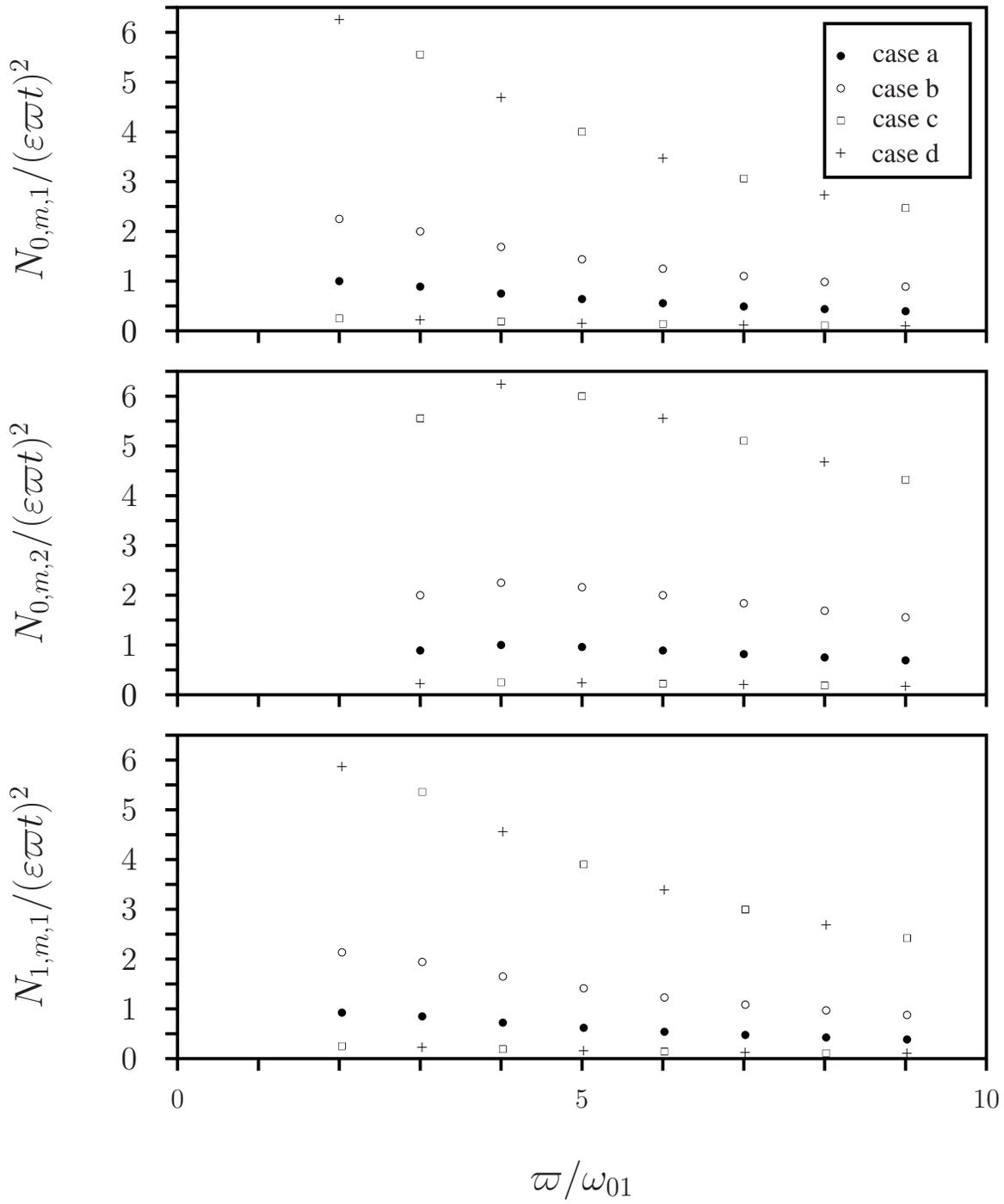
\end{document}